\documentclass[twocolumn,prl]{revtex4}

\def\gev{\mbox{GeV}}
\def\mev{\mbox{MeV}}

\def\cm{\mbox{cm}}
\def\mpc{\mbox{Mpc}}
\def\AJ{{\it Ap. J.} }
\def\AJL{{\it Ap. J. Lett.} }

\def\APJ{{\it Ap. J.} }

\def\IJMP{{\it Int. J. Mod. Phys.} }

\def\MNRAS{{\it Mon. Not. R. Ast. Soc.} }
\def\NAT{{\it Nature} }
\def\NAST{{\it New Astronamy}} 

\def\NP{{\it Nucl. Phys.} }
\def\PL{{\it Phys. Lett.} }

\def\PRL{{\it Phys. Rev. Lett.} }

\def\SC{{\it Science} }

\def\vev#1{\langle {#1}\rangle}

\def\frac#1#2{{\textstyle{{#1}\over {#2}}}}

\def\lsim{\mathrel{\rlap{\lower4pt\hbox{\hskip1pt$\sim$}}
    \raise1pt\hbox{$<$}}}
\def\gsim{\mathrel{\rlap{\lower4pt\hbox{\hskip1pt$\sim$}}
    \raise1pt\hbox{$>$}}}
\def\sqr#1#2{{\vcenter{\vbox{\hrule height.#2pt
         \hbox{\vrule width.#2pt height#1pt \kern#1pt
         \vrule width.#2pt}
         \hrule height.#2pt}}}}

\newcommand{\beq}{\begin{equation}}
\newcommand{\eeq}{\end{equation}}
\newcommand{\bea}{\begin{eqnarray}}
\newcommand{\eea}{\end{eqnarray}}

\begin{document}

\vbox to0mm{\vspace{-10.mm}  \hbox to18cm {\hfill \mbox{DF/IST-1.2000;\,
IFT-P.028/2000}}}
	
\title{Self-interacting Dark Matter and Invisibly Decaying Higgs}

\author{M.C. Bento}
\author{O. Bertolami}
\email{orfeu@cosmos.ist.utl.pt}
\affiliation{Instituto Superior T\'ecnico, Departamento de F\'\i sica, 
Av. Rovisco Pais 1, 1049-001 Lisboa, Portugal}
\author{R. Rosenfeld}
\email{rosenfel@ift.unesp.br}
\affiliation{Instituto de F\'\i sica Te\'orica, R.\ Pamplona 145, 
01405-900 S\~ao Paulo - SP, Brazil }
\author{L. Teodoro}
\email{lteodoro@glencoe.ist.utl.pt}
\affiliation{Centro Multidisciplinar de Astrof\'\i sica, 
Instituto Superior T\'ecnico, Av. Rovisco Pais 1, 1049-001 Lisboa, Portugal}

\date{\today}

\begin{abstract}
Self-interacting dark matter has been suggested in order to
overcome the difficulties of the Cold Dark Matter model on
galactic scales. We argue that a scalar gauge singlet coupled to
the Higgs boson, which could lead to an invisibly decaying Higgs,
is an interesting candidate for this self-interacting dark matter
particle. We also present estimates on the abundance of these
particles today as well as consequences to non-Newtonian forces.
\end{abstract}

\maketitle

\section{Introduction}
\label{int}

Finding clues for the nature of dark matter (DM) in the Universe is 
one of the most
pressing issues in the interface between particle physics and cosmology.
The cold dark matter model supplemented by a cosmological constant
($\Lambda$CDM), in the context of inflationary models, explains 
successfully the observed structure of the Universe on large scales,
the cosmic microwave background anisotropies and  type Ia 
supernovae observations \cite{Bahcall} for a given set of density parameters,
{\it e.g.},  $\Omega_{DM} \sim 0.30$, 
$\Omega_{Baryons} \sim 0.05$ and $\Omega_{\Lambda} \sim 0.65$. According to 
this scenario, initial
Gaussian density fluctuations, mostly in non-relativistic collisionless
particles, the so-called cold dark matter, are generated in an inflationary 
period of the Universe. These fluctuations grow gravitationally forming dark
halos into which luminous matter is eventually condensed and cooled. 

However, despite its successes, there is a growing wealth
of observational data that raise problems in
the CDM scenarios. N-body simulations predict a number of halos 
which is a factor $\sim$ 10 larger than the 
observed number at the 
level of Local Group \cite{Mooreetal2,Klypin}. Furthermore, CDM models 
yield dispersion velocities in the Hubble flow within a sphere 
of $5~h^{-1}~$Mpc 
between $300-700~$kms$^{-1}$ for $\Omega_{DM} \sim 0.95$ and 
between $150-300~$kms$^{-1}$  for $\Omega_{DM} \sim 0.30$. The observed 
value is about 
$60~$kms$^{-1}$. Neither model 
can produce a single  Local Group candidate with the observed velocity 
dispersion in a volume of 
$10^{6}~h^{-3}\mbox{Mpc}^{3}$ \cite{Governato}. A 
related issue is that astrophysical systems which are DM dominated like 
the core, 
dwarf galaxies \cite{Moore,Flores-Primack,Burkert1}, low surface brightness 
galaxies
\cite{Blok} and galaxy clusters without a central cD galaxy \cite{Tyson}
show shallow matter--density profiles which can be modeled by 
isothermal spheres with finite central densities.
This is in contrast with galactic and galaxy cluster halos in high 
resolution N-body simulations 
\cite{Navarro,Ghigna0,Ghigna,Mooreetal} which have singular cores, 
with $\rho \sim r^{-\gamma}$ and $\gamma$ in the range between 1 and 2.
Indeed, cold collisionless DM particles do not have 
any associated length scale leading, due to hierarchical gravitational 
collapse, to dense dark matter halos with negligible 
core radius \cite{Hogan}.

It has been argued that astrophysical processes such as feedback from  
star formation or an ionizing background to inhibit star formation and
expelling gas in low mass halos \cite{Dekel,Quinn,Navarro2} may solve 
some of abovementioned problems. However, such processes have 
been difficult to accommodate in our 
understanding of galaxy formation since galaxies outside clusters 
are predominantely rotationaly supported 
disks and their final structure does not result from  the struggle 
between gravity and winds but rather are set by their initial angular momentum.

Another possible solution, coming from particle physics, would be to 
allow DM particles to self-interact so that they have a large scattering 
cross section and negligible annihilation or dissipation. The self-interaction
results in a characteristic length scale given by the mean free 
path of the particle in the halo.
This idea has been originally proposed to suppress small scale power 
in the standard CDM model  
\cite{Carlson,Laix} and has been recently revived in order to address the
issues discussed above \cite{Spergel}. The main feature of self-interacting
dark matter (SIDM) is that large self-interacting cross sections lead to a
short mean free path, so that dark matter particles with mean free 
path of the order of the scale length of halos allows for the transfer of 
conductive heat to the halo cores, a quite desirable feature \cite{Spergel}. 
Recently performed numerical simulations indicate that strongly 
self-interacting dark matter does indeed lead to better predictions 
concerning satellite 
galaxies \cite{Hannestad,Moore1,Yoshida}. However, only in presence  
of weak self-interaction \cite{Burkert} the core problem might be solved.

The two-body cross section is estimated to be in the range of $\sigma/m \sim 
10^{-24} $ to $10^{-21}$ cm$^2$/GeV, from a variety of arguments, including 
a mean free path between 1 and 1000 kpc \cite{Spergel}, requiring  the core
expansion time scale to be smaller than the halo age \cite{Burkert1,Firmani}
and analysis of cluster ellipticity \cite{Escude}. A larger value of $\sigma/m
\sim  10^{-19} $ cm$^2$/GeV was obtained from a best fit to the rotation curve 
of a low surface brightness in a simulation where some extra simplying
assumptions were made \cite{Hannestad}. In this work, we shall assume 
for definiteness that the
the cross section is fixed via the requirement  that the mean free path of the
particle in the halo is in the range  $1-1000$ kpc.

\section{A model for self-interacting, non-dissipative CDM}
\label{model}

Many models of physics beyond the Standard Model suggest the existence of new
scalar gauge singlets, {\it e.g.}, in the so-called next-to-minimal
supersymmetric standard model \cite{NMSSM}. 
In this section, we provide a simple example for the realization of the 
idea proposed in \cite{Spergel} of a self-interacting, non-dissipative cold 
dark matter candidate that is based on an extra gauge singlet, 
$\phi$, coupled to the standard model Higgs boson, $h$, with a Lagrangian 
density given by:
\begin{equation}
{\cal L} = {1 \over 2} (\partial_\mu \phi)^2 - {1 \over 2} m_\phi^2 \phi^2  
- {g \over 4} \phi^4 + g' v \phi^2 h ~,
\end{equation}%
where $g$ is the field $\phi$
self-coupling constant, $m_\phi$ is its mass, $v=246$ GeV is the Higgs 
vacuum expectation value and $g'$ is the coupling between the singlet
$\phi$ and  $h$. We assume that the $\phi$ mass does not arise from
spontaneous symmetry breaking since, as we shall see in the next section, tight
constraints from non-Newtonian forces eliminates this possibility due to the
fact that, in this case, there is a relation among coupling constant, mass and
vacuum expectation value that results in a tiny scalar self-coupling constant.
In its essential features our self-interacting dark matter model can be 
regarded as a concrete realization of the generic massive scalar field with 
quartic potential discussed in \cite{Peebles,Goodman}.

We shall assume that $\phi$ interacts only with $h$ and with
itself. It is completely decoupled for $g' \rightarrow 0$.  
For reasonable values of $g'$, this new scalar would introduce a new, invisible
decay mode for the Higgs boson. This could be an important loophole in the
current attemps to find the Higgs boson at accelerators \cite{Bij}.
This coupling could, in principle, be relevant for $\phi \phi$ scattering but
we shall be conservative and assume that it is small and neglect 
its contribution. However, we point out that even for non-zero values of $g'$, 
the new scalar is stable in this model.

These particles are non-relativistic, with typical velocities of 
$v \simeq 200 $km $s^{-1}$. 
Therefore, it is not possible to dissipate energy by, for instance,
creating more particles in reactions like $\phi \phi \rightarrow \phi \phi \phi
\phi$. Only the elastic channel is kinetically accessible and the scattering 
matrix element near threshold ($s \simeq 4 m_\phi^2$) is given by:
\begin{equation}
{\cal M} (\phi \phi \rightarrow \phi \phi) = i g~~. 
\end{equation}
Near threshold the cross section is given roughly by:
\begin{equation}
\sigma (\phi \phi \rightarrow \phi \phi) \equiv \sigma_{\phi \phi} 
= {g^2 \over 16 \pi s} \simeq {g^2 \over 64 \pi m_\phi^2} ~~.
\label{cross1} 
\end{equation}

We shall derive limits on $m_\phi$ and $g$ by demanding that the mean free path
of the particle $\phi$, $\lambda_\phi$, should be in the interval 
$1~\mbox{kpc} < \lambda_{\phi} < 1~$Mpc. This comes about because, 
if the mean free path  were much greater than about $1~$Mpc, dark 
matter particles would not experience any interaction as they fly through a
halo. On the other hand, if the dark matter mean free path were much
smaller than $1~$kpc, dark matter particles would behave as a collisional  
gas altering substantially the halo structure and evolution. 
Hence, we have:
\begin{equation}
\lambda_{\phi} = {1 \over \sigma_{\phi \phi} n_\phi} 
= {m_\phi \over \sigma_{\phi\phi} \rho^{h}_\phi} ~~, 
\end{equation}
where $n_\phi$ and $\rho^{h}_\phi$ are the number and mass density in 
the halo of the $\phi$ particle, respectively. 
Using $\rho^{h}_\phi = 0.4$ GeV/cm$^3$, 
corresponding to the halo density, one finds:
\begin{equation}
\sigma_{\phi \phi} = 2.1 \times 10^{3}~ 
\left({m_\phi \over \gev}\right) 
\left({\lambda_\phi \over \mpc} \right)^{-1}~ \gev^{-2} ~~.
\label{cross2}
\end{equation} 

Equating Eqs. (\ref{cross1}) and (\ref{cross2}) we obtain:
\begin{equation}
m_\phi= 13~g^{2/3}
\left({\lambda_{\phi} \over \mpc} \right)^{1/3} \mev ~~.
\label{mass}
\end{equation}

Demanding the mean free path of the $\phi$
particle to be of order of 1 Mpc implies in the {\it model independent} 
result:
\begin{equation}
{\sigma_{\phi \phi} \over m_\phi} = 8.1 \times 10^{-25} 
\left({\lambda_{\phi} \over \mpc} \right)^{-1} \cm^2/\gev ~~.
\label{SStein}
\end{equation}

Recently, it has been argued, on the basis of  gravitational lensing analysis, 
that the shape of the MS2137 - 23 system is elliptical while 
self-interacting non-dissipative CDM implies that halos are 
spherical \cite{Escude}. Furthermore, the limit
\begin{equation}
{\sigma_{\phi \phi} \over m_\phi} < 10^{-25.5} ~\cm^2/\gev 
\label{Escude}
\end{equation}%
arises from that analysis, which is about an order of magnitude smaller 
than (\ref{SStein}). Indeed, gravitational lensing arguments are acknowledged
to be crucial in validating SIDM; however, estimates made in \cite{Escude}
were criticized as they rely on a single system and because their intrinsic 
uncertainties actually allow for consistency with SIDM \cite{Moore1}. 

Let us now estimate the amount of $\phi$ particles that were produced in the 
early Universe and survived until present. We shall assume that $\phi$ 
particles were mainly produced during reheating after the end of inflation. A 
natural setting to consider this issue is within the framework of 
${\cal N} = 1$ supergravity
inspired inflationary models where the inflaton sector couples with the 
gauge sector through the gravitational interaction. Hence, 
the number of $\phi$ particles expressed in terms of the ratio 
$Y_{\phi} \equiv {n_{\phi} \over s_{\gamma}}$, where $s_{\gamma}$ is the 
photonic entropy density, is related with the inflaton ($\chi$) 
abundance after its decay by 
\begin{equation}
Y_{\phi} = {1 \over N} Y_{\chi} ~~, 
\label{inflaton}
\end{equation}%
where $N$ is the number of degrees of freedom. Notice that $Y_\phi$ 
is a conserved quantity since $\phi$ does not couple to fermions. 
In the context of ${\cal N} = 1$ supergravity inflationary models, 
given the upper bound on the 
reheating temperature in order to avoid the gravitino problem (see \cite{Bento}
and references therein), $Y_{\chi}$ is given by the ratio of the reheating 
temperature and the inflaton mass and, for typical models%
\begin{equation}
Y_{\chi} = {T_{RH} \over m_{\chi}} = \epsilon~10^{-4} ~~, 
\label{reheating}
\end{equation}%
where $\epsilon$ is an order one constant. This estimate allows us to compute
the energy density contribution of $\phi$ particles in terms of the baryonic
density parameter:%
\begin{equation}
\Omega_{\phi} = {1 \over N} {T_{RH} \over m_{\chi}} {1 \over \eta_{B}} 
{m_{\phi} \over m_{B}}~\Omega_{B}~~, 
\label{density}
\end{equation}%
where $\eta_{B} \simeq 5 \times 10^{-10}$ is the baryon asymmetry of the 
Universe \cite{Olive}.

Using Eq. (\ref{mass}) and taking $N \simeq 150$, we obtain: 

\begin{equation}
\Omega_{\phi} \simeq 18.5~\epsilon~g^{2/3}
\left({\lambda_{\phi} \over \mpc} \right)^{1/3} \Omega_{B} ~~, 
\label{finaldensity}
\end{equation}
which allows identifying $\phi$ as the cosmological dark matter candidate,
i.e. $\Omega_{\phi} \simeq \Omega_{DM} \lsim 0.3 $ \cite{Bahcall1},
for $\epsilon \sim 0.5$, $g$ of order one and $\lambda_{\phi}$ of 
about $1~$Mpc.

\section{The case of a non-Newtonian interaction}

In the previous section we have considered self-interacting 
DM particles interacting with ordinary matter via gravity. In this 
section, we shall consider the possibility of allowing the self-interacting
DM particle to couple to ordinary matter via a non-Newtonian type 
force as well. This possibility has been intensively 
discussed in the past 
and repeatedly sought in laboratory (see \cite{Bertolami} and references 
therein). Moreover, 
it has recently been revived in the context of the accelerated 
expansion of the Universe \cite{Fujii}. This is a fairly interesting 
possibility as 
the carrier of a putative new interaction can quite naturally be regarded as a 
DM candidate \cite{Bertolami,Gradwohl,Stubbs}. 
In what follows, we will show that the
carrier of the non-Newtonian force necessarily has an extremely small 
self-coupling. 

Assuming that the Lagrangian density of the new force carrier, $\varphi$, 
is given by%
\begin{equation}
{\cal L} = {1 \over 2} (\partial_\mu \varphi)^2 - {1 \over 2} m_{\varphi}^2 
\varphi^2  - {g_{\varphi} \over 4} \varphi^4 ~,
\label{eq:3.1}
\end{equation}%
while its coupling with nucleons of mass, $m_{N}$, and photons is given by
\begin{equation}
{\cal L}_{int} = c_{N} {\varphi \over \vev{\varphi}} m_{N} \psi \bar \psi 
+ c_{G} {\varphi 
\over \vev{\varphi}} F_{\mu \nu} F^{\mu \nu}~~,   
\label{eq:3.2}
\end{equation}
where $\vev{\varphi}$ is a large scale associated to the new interaction and
$c_{N}$, $c_{G}$ are coupling constants. This last interaction implies that
$\varphi$ exchange leads to a non-Newtonian contribution for the 
interaction energy, $V(r)$, between two 
point masses $m_1$ and $m_2$, that can be expressed in terms of 
the gravitational interaction as%
\begin{equation}
V(r) = -  {G_{\infty}~m_{1}~m_{2} \over r}~(1 + \alpha_5~e^{-r/\lambda_5})~,
\label{eq:3.3}
\end{equation}
where $r = \vert \vec r_{2} - \vec r_{1} \vert$ is the distance between the
masses, $G_{\infty}$ is the gravitational coupling for $r \rightarrow \infty$, 
$\alpha_5$ and $\lambda_5$ are the strength and the range of the new 
interaction so that $\lambda_5 = m_{\varphi}^{-1}$ and%
\begin{equation}
\alpha_5 = {c_{N}^2 \over 4 \pi} \left({M_{p} \over \vev{\varphi}}\right)^2~~, 
\label{eq:3.4}
\end{equation}  
where $M_P \equiv G_{\infty}^{-1/2}$ is the Planck mass. Existing bounds on
$\alpha_5$ (see \cite{Bertolami}) 
imply for $c_{N}$ of order one that $\vev{\varphi} \sim M_P$. 
If however, $c_{N} \lsim 10^{-5}$, than one could have instead 
$\vev{\varphi} \sim 10^{-5}~M_P$ \cite{Hill}. 

The issue is, however, that in order to generate a vacuum expectation value to
$\varphi$, from Eq. (\ref{eq:3.1}) so as to satisfy Eq. (\ref{mass}), 
one must have
$m_{\varphi}^2 < 0$, from which would imply that 
$g_{\varphi} = \left({\vert m_{\varphi} \vert \over \vev{\varphi}}\right)^2 
<< 1$, meaning that 
for either choice of $\vev{\varphi}$ quoted above, $\varphi$ DM particles 
have a negligible self-interaction.
This argument can be generalized for 
any potential that gives origin to a vacuum expectation value for $\varphi$
as specified above. We can therefore conclude that the carrier of 
a non-Newtonian force is not an acceptable candidate for SIDM.

\vskip 0.3cm

\section{Outlook}

In this letter we suggest that a scalar gauge singlet coupled with the Higgs
field in a way to give origin to an invisible Higgs is a suitable candidate
for self-interacting dark matter. This proposal has some distinct features. 
Firstly, since gauge invariance 
prevents the scalar singlet to couple to fermions, hence 
strategies for directly searching this dark matter candidate must necessarily 
concentrate on the hunt of the Higgs field itself in accelerators. 
Furthermore, in what concerns its astrophysical and cosmological implications
the relevant features of our proposal are quite unambiguously expressed 
by Eqs. (\ref{cross2}), (\ref{mass}) and (\ref{finaldensity}). 
Confronting the result of simulations with our candidate for different 
values of the relevant parameters with observations may turn out to 
be crucial in validating our proposal.  

\vskip 0.3cm

{\bf Acknowledgments} 

\vskip 0.3cm

\noindent
One of us (O.B.) would like to thank the hospitality of
Instituto de F\'\i sica Te\'orica in S\~ao Paulo, Brazil, where
part of this work was carried out. R.R. is supported by CNPq
(Brazil) and FAPESP (S\~ao Paulo). L.T. would like to acknowledge
the financial support from Funda\c c\~ao para a Ci\^encia e a
Tecnologia (Portugal) under the grant PRAXIS XXI /BPD/16354/98 as well as
the project PRAXIS/C/FIS/13196/98, and Ben Moore for pointing out
the astrophysical consequences of self-interacting dark matter.

\end{document}